\documentclass[aps,prl,longbibliography,twocolumn,superscriptaddress]{revtex4-1}

\usepackage[pdftex]{graphicx}
\usepackage{amsmath}
\usepackage{amssymb}
\usepackage{bm}
\usepackage[normalem]{ulem}
\usepackage{enumerate}
\usepackage{mathtools}
\usepackage[dvipsnames]{xcolor}
\usepackage[pdftex]{hyperref}
\hypersetup{
	colorlinks=true,
	linkcolor=MidnightBlue,
	citecolor=MidnightBlue,
	urlcolor=MidnightBlue
}

\let\oldepsilon\epsilon
\let\epsilon\varepsilon
\let\varepsilon\oldepsilon

\def\0v{\mathbf{0}}

\def\rv{\mathbf{r}}

\begin{document}

\title{{Intermittent Attractive Interactions Lead to Microphase Separation in Non-motile Active Matter}}

\author{Henry Alston}
\affiliation{Department of Mathematics, Imperial College London, 180 Queen’s Gate, London SW7 2BZ, United Kingdom}
\author{Andrew O. Parry}
\affiliation{Department of Mathematics, Imperial College London, 180 Queen’s Gate, London SW7 2BZ, United Kingdom}
\author{Rapha\"el Voituriez}
\affiliation{Laboratoire de Physique Th\'eorique de la Mati\`ere Condens\'ee, UMR 7600 CNRS/UPMC, 4 Place Jussieu, 75255 Paris Cedex, France}
\affiliation{Laboratoire Jean Perrin, UMR 8237 CNRS/UPMC, 4 Place Jussieu, 75255 Paris Cedex, France}
\author{Thibault Bertrand}
\email{t.bertrand@imperial.ac.uk}
\affiliation{Department of Mathematics, Imperial College London, 180 Queen’s Gate, London SW7 2BZ, United Kingdom}

\date{\today}

\begin{abstract}
Non-motile active matter exhibits a wide range of non-equilibrium collective phenomena yet examples are crucially lacking in the literature. We present a microscopic model inspired by the bacteria \textit{Neisseria Meningitidis} in which diffusive agents feel intermittent attractive forces. Through a formal coarse-graining procedure, we show that this truly scalar model of active matter exhibits the time-reversal-symmetry breaking terms defining the \textit{Active Model B+} class. In particular, we confirm the presence of microphase separation by solving the kinetic equations numerically. We show that the switching rate controlling the interactions provides a regulation mechanism tuning the typical cluster size, e.g. in populations of bacteria interacting via type IV pili.
\end{abstract}


\maketitle

All matter is built up from smaller components, active matter is no different. Often of biological inspiration, active matter generically denotes systems of particles which consume energy from their surroundings \cite{Ramaswamy2010,Gompper2020}. While this continuous consumption of energy leads to the breaking of time-reversal symmetry (TRS) at the microscopic scale and thus maintains active systems out of equilibrium, striking non-equilibrium features generically stem from interactions between active particles or with their environment \cite{Marchetti2013,Bechinger2016,Bertrand2018}. For instance, dense suspensions of interacting self-propelled particles display a wealth of phenomena forbidden by equilibrium thermodynamics including long-range order \cite{Vicsek1995,Deseigne2010,Bricard2013,Solon2015,Bertrand2020b}, clustering \cite{Theurkauff2012,Buttinoni2013,Palacci2013,Bertrand2020a} or phase separation {even} in the absence of attractive interactions (e.g. motility induced phase separation) \cite{Bialke2013, Speck2014, Speck2015,Zhang2021,Tailleur2008,Fily2012,Redner2013}. Connecting emergent structures and collective dynamics to {the behavior of individual particles} through coarse-graining techniques remains an open problem which has seen recent development \cite{Lowen2020,Cates2015,Farrell2012,Bertin2009,Dean1996}.

Equilibrium phase separation remains one of the simplest examples of order emerging from disorder, characterized by the spontaneous formation of regions with contrasting characteristics within a system. The dynamics of phase separation in a passive binary fluid are captured by Halperin and Hohenberg's \textit{Model B} \cite{Hohenberg1977} which describes the evolution of a conserved scalar order parameter in a system {respecting} time reversal symmetry (TRS) \cite{Bray2002,Barrat2003,Lowen1994,Chaikin1995}. {\it Model B} itself can be derived from \textit{Dynamical Density Functional Theory} --- central to the analysis of passive, soft matter systems \cite{TeVrugt2020a, Archer2004}.

In contrast, recent works have focused on field theories capturing the TRS breaking present in active systems. Using a top-down approach, TRS violating terms can be added to {\it Model B} equations to form a mean-field theory for motility induced phase separation leading to the so-called {\it Active Model B} \cite{Wittkowski2014}. Interestingly, the addition in this active field theory of further terms (of the same order in the expansion in the order parameter) leads to a non-equilibrium field theory, {\it Active Model B+} (AMB+), which has been shown numerically and analytically to display microphase separation, driven by a \textit{reverse} Ostwald ripening (ROR) process \cite{Tjhung2018,Fausti2021}. The suppression of Ostwald ripening was also discussed in the context of coarse-grained models of \textit{active emulsions} used to study phase separation in systems driven out-of-equilibrium, e.g. by chemical reactions \cite{Zwicker2014, Zwicker2015, Wurtz2018,Weber2019}. 

In many-body physics, complex and robust collective behaviors can be the result of interactions between very simple constituent agents. While previous coarse-graining approaches have successfully produced the AMB+ equation, these bottom-up approaches have focused on {\it motile} active matter {---} by far the most studied class of active systems. In contrast, minimal models of non-motile {--- and in a sense {\it truly} scalar ---} active matter are crucially lacking in the literature, although they offer further examples of the non-equilibrium phenomena present in biological systems. Breaking from {the} motile active matter paradigm, we introduce in this Letter a minimal microscopic model of particles whose interactions are governed by an active stochastic process.

{\it Active switching} was previously introduced in microscopic models to generate particle shape changes \cite{Grawitter2018}, define the particle-particle interactions \cite{Bonazzi2018,MonchoJorda2020,Bley2021} or particles interactions with an external field \cite{Zakine2018}. Our model is inspired by the bacterium \textit{Neisseria Meningitidis} which interacts with its neighbours and environment through type IV pili, hair-like appendages whose contraction generates pulling forces \cite{Bonazzi2018,Kuan2021}. In isolation, the bacterium extends and retracts its pili over time. Upon proliferation, the pili of neighbouring bacteria touch; following contact, their retraction pulls pairs of bacteria together, eventually leading to bacterial clustering.

Recently,  the mechanical properties of bacterial aggregates were explored using experiments and phenomenological continuum models \cite{Kuan2021,Zhou2021,Oriola2021}. In contrast, we describe minimally the pili interaction and introduce a model in which particles stochastically switch between attractive and purely repulsive states. We argue that this effective description loses none of the fundamental physics but allows for significant analytical progress. While the symmetries of our microscopic model are consistent with \textit{Active Model B} and \textit{B+}, a formal coarse-graining is required to conclude. We derive a density equation which we show is of AMB+ form by identifying the TRS breaking terms \cite{Tjhung2018, Fausti2021}. Finally, we confirm the presence of microphase separation and reverse Ostwald Ripening as predicted by the field theory by solving the kinetic equations numerically and compare these results to direct numerical simulations of the microscopic model, fully characterising the non-equilibrium structure displayed by the system.

\begin{figure}[t!]
\centering
\includegraphics[width=85mm]{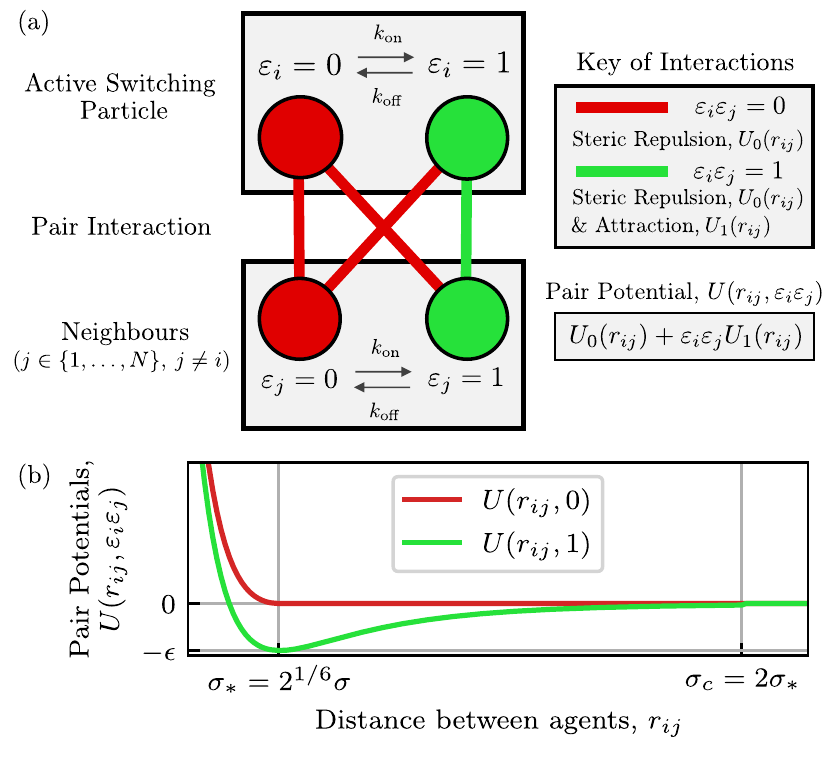}
\caption{\textit{Schematic of Microscopic Interactions.} (a) The state of particle $i$ is set by its internal variable, $\epsilon_i$, which switches between $0$ and $1$ with fixed rates, $k_{\rm{on}}$ and $k_{\rm{off}}$. The pair potential for neighboring particles depends on the product $\epsilon_i\epsilon_j$. (b) Pair potentials used in the simulations for $\epsilon_i\epsilon_j=0$ (red) and $\epsilon_i\epsilon_j=1$ (green). A WCA potential sets the particle size $\sigma_{*}$ and the attraction range is set to $\sigma_c = 2\sigma_*$ \cite{Note1}.}
\label{fig:micromodel}
\end{figure}

\textit{Microscopic model ---}
We consider a system of $N$ particles characterised by their position $\mathbf{r}_i$ and an internal variable $\epsilon_i\in\{0,1\}$ defining their interactions. Any two particles interact through steric repulsion when their center-to-center distance is such that $|\mathbf{r}_i-\mathbf{r}_j|=r_{ij}<\sigma_*$, independently of the value of $\epsilon_i$ and $\epsilon_j$. If the internal variables of both agents are such that $\epsilon_i=\epsilon_j=1$, these particles are {additionally} subjected to an attractive force with longer range $\sigma_c > \sigma_*$ (see Fig.\,\ref{fig:micromodel}(a)). We refer to the case where $\epsilon_i = 1$ (resp., $\epsilon_i = 0$) as the {\it on} state (resp., the {\it off} state). We can define the total {pair} interaction potential as the superposition of purely repulsive $U_0$ and purely attractive $U_1$ contributions (see Fig.\,\ref{fig:micromodel}(b)):
\begin{equation}
U(r_{ij}, \epsilon_i\epsilon_j)=U_0(r_{ij})+ \epsilon_i\epsilon_j U_1(r_{ij}).
\label{eq:fullpotential}
\end{equation}

The motion of the particles is governed by the overdamped Langevin equation
\begin{equation}
\dot{\mathbf{r}}_i = -\frac{1}{\gamma}\sum_{j\ne i} \nabla_{\mathbf{r}_i} U(r_{ij}, \epsilon_i\epsilon_j) + \sqrt{2D}\bm{\eta}_i,
\label{eq:eom}
\end{equation}
where $\gamma$ is a friction coefficient, $D$ is the bare-diffusion coefficient which sets the temperature in the system and $\bm{\eta}_i$ is a zero mean, unit variance Gaussian white noise.

We introduce activity by allowing the particles to stochastically switch between the {\it on} and {\it off} states with constant rates, generically leading to intermittent attractive forces (Fig.\,\ref{fig:micromodel}(a)). Formally, the internal variables $\{\epsilon_i\}_{i\in[1,N]}$ follow independent telegraph processes \cite{VanKampen2007} with switching rates $k_{\rm on}$ and $k_{\rm off}$ (see \footnote{See Supplemental Material at []} for details). 

\begin{figure}[t!]
	\centering
	\includegraphics[width=0.45\textwidth]{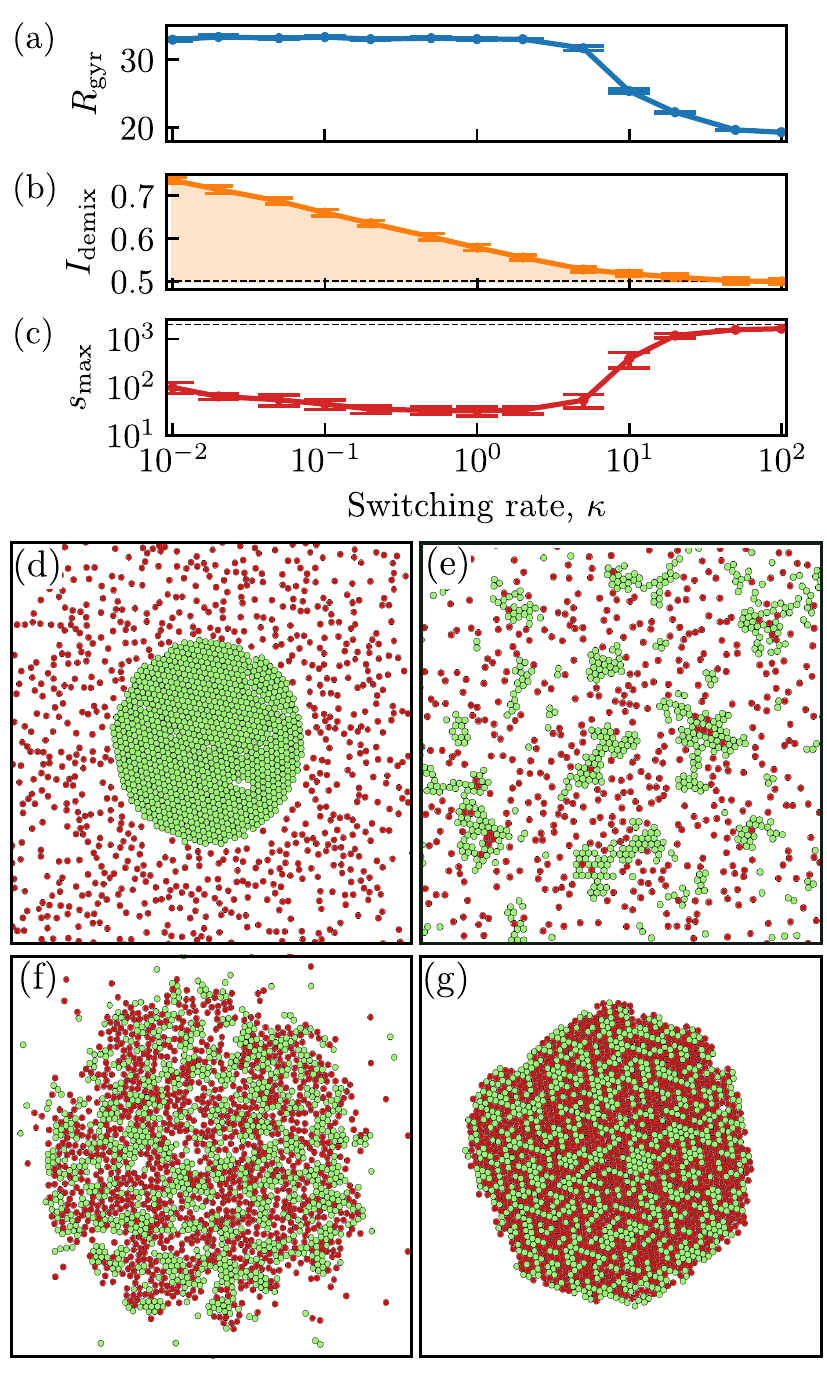}
	\caption{\textit{Emergent Structures in Active Switching System.} (a) Radius of gyration $R_{\rm gyr}$, (b) demixing index $I_{\rm demix}$ and (c) maximal cluster size $s_{\rm max}$ for switching rates $\kappa \in [10^{-2},10^2]$ and $\varepsilon\gg k_BT$.  Representative configurations obtained in simulations at steady-state for (d) $\kappa=0$, (e) $\kappa=10^{-2}$, (f) $\kappa=10$, and (g) $\kappa=100$.}
	\label{fig:emergentstructure}
\end{figure}

\textit{Microscopic simulations ---} 
First, we numerically solve the equation of motion \cite{Redner2013, Branka1999}. The interaction potentials $U_0$ and $U_1$ are defined following the WCA decomposition \cite{Note1,Weeks1971}. We believe our results to be insensitive to the exact choice of potential. To ensure that the system exhibits liquid-gas phase separation with no active switching, we work in the limit $\varepsilon  \gg k_B T$. Here, we restrict our focus to the case where $k_{\rm{on}}=k_{\rm{off}} = k$. We non-dimensionalize the switching rate setting $\kappa=k\sigma^2/D$, where $\sigma$ is the nominal particle diameter.

As we vary the switching rate between {$10^{-2} \le \kappa \le 10^2$}, we investigate the emergence of macroscopic structures (Fig.\,\ref{fig:emergentstructure}). At large switching rates $\kappa \gg 1$, the system fully phase separates and displays a single macroscopic drop as can be seen on Fig.\,\ref{fig:emergentstructure}(f)-(g); this is evidenced by a low radius of gyration $R_{\rm gyr}$ for $\kappa\geq10$ as well as a maximal cluster size $s_{\rm max}$ approaching the system size. Further, we observe that the stable drop is fully mixed with a demixing index $I_{\rm demix} \approx 0.5$, defined as the fraction of neighboring particles in the same state \cite{Note1}. For $\kappa\gg1$, the diffusion timescale is much larger than the time between switching events; agents do not have time to diffuse out of reach of the central drop before switching \textit{on} and being pulled back.

As the switching rate decreases, both radius of gyration and demixing index monotonically increase. At low switching rates, the system does not reach full phase separation; instead, we argue that at intermediate switching rates our model exhibits \textit{microphase} separation, where the system supports the coexistence of a large number of small clusters (Fig.\,\ref{fig:emergentstructure}(e)). We conclude that the system demixes and self-organizes into clusters of {\it on} particles surrounded by a gas of {\it off} particles. The maximal cluster size reaches a minimum when $\kappa \approx 1$ and increases again as we lower $\kappa$. Liquid-gas phase separation and demixing are strengthened as $\kappa$ decreases. Indeed, longer times between switching events allow the nucleated clusters of attractive particles to grow further. In the singular limit where $\kappa=0$, we observe a fully demixed state displaying a stable single drop of {\it on} particles surrounded by a diffusive gas of {\it off} particles. In the limit $\varepsilon  \gg k_B T$, the initial fraction of {\it on} particles controls the size of this drop as strong attraction ensures that {\it on} particles remain in the condensed phase.

\textit{Kinetic equations ---}
Starting from a many-body Smoluchowski equation and solving the subsequent BBGKY hierarchy, we explicitly coarse-grain our microscopic model (see details in SM \cite{Note1}). Such a derivation is often omitted for active switching systems, instead these kinetic equations form the basis of the \textit{Reaction-Diffusion DFT} (R-DDFT) framework \cite{MonchoJorda2020, Grawitter2018, Zakine2018, TeVrugt2020a, TeVrugt2020b}.

At a macroscopic level, we find that the state of our system is described by the density fields $\rho_0(\mathbf{r}, t)$ and $\rho_1(\mathbf{r}, t)$ for the {\it off} and {\it on} particles, respectively, which are governed by the following kinetic equations
\begin{subequations}
\begin{align}
	\partial_t\rho_0(\rv, t) &= \nabla\cdot \mathbf{J}_0 + s(\rho_0, \rho_1) \label{eq:kineq1}\\
	\partial_t\rho_1(\rv, t) &= \nabla\cdot \mathbf{J}_1 - s(\rho_0, \rho_1) \label{eq:kineq2}
\end{align}
\label{eq:kineticequations}%
\end{subequations}
where the effect of the active switching is entirely contained in the coupling term $s(\rho_0, \rho_1) =  k( \rho_1 - \rho_0)$. Self-diffusion and particle-particle interactions are expressed through the fluxes: 
\begin{subequations}
\begin{align}
\mathbf{J}_0 &= D\nabla\rho_0 + \rho_0\nabla\mu_{\rm{rep}}\big(\rho\big) \label{eq:current1} \\
\mathbf{J}_1 &= D\nabla\rho_1 + \rho_1\nabla\mu_{\rm{rep}}\big(\rho\big)+\frac{1}{2}\rho_1\nabla\big(U_1\star \rho_1\big) \label{eq:current2}
\end{align}
\label{eq:currents}%
\end{subequations}
where $\rho(\rv, t)=\rho_0(\rv, t)+\rho_1(\rv, t)$ is the total particle density. We note that although both {\it on} and {\it off} particles are subject to steric interactions, only {\it on} particles are subject to attractive interactions (last term in Eq.\,(\ref{eq:current2})). 

Interestingly, we note that in the case where $k=0$, equations (\ref{eq:kineticequations}) and (\ref{eq:currents}) describe two classical equilibrium systems: a hard-sphere gas and a phase-separating Cahn-Hilliard-type fluid. Our results so far show that by coupling these two fluids, the resulting system can exhibit fundamentally non-equilibrium phase separation behaviours, including microphase separation. While this has been hinted at in previous studies of \textit{Active Emulsions} using phenomenological continuum models \cite{Zwicker2015, Wurtz2018,Zwicker2014,Weber2019}, we here derive a closed equation for $\rho(\rv,t)$ and show formally that it pertains to the AMB+ class. 

\textit{Closed Equation for Agent Density ---} Starting from Eq.\,(\ref{eq:kineticequations}), we write an equation for the total density of particles
\begin{equation}
\partial_t \rho(\rv, t)=\nabla\cdot\bigg[\rho(\rv)\nabla\bigg(\frac{\delta\mathcal{F}[\rho(\rv)]}{\delta\rho(\rv)}\bigg)+\frac{1}{2}{\rho_1(\rv)}\nabla\big(U_1\star \rho_1\big)\bigg]
\label{eq:rho_open}
\end{equation}
where although one cannot generically write a free energy functional for active systems, we follow a common notation in field theories of active phase separation \cite{Wittkowski2014,Tjhung2018,Fausti2021} and write the passive terms in our density equation as the gradient of the functional derivative of a free energy-like functional
\begin{equation}
\mathcal{F}[\rho(\rv)] = \int d\rv\:\big[ D\rho(\rv)\big[\log(\rho(\rv))-1\big] + f_{\rm{rep}}(\rho(\rv))\big].
\label{eq:free_energy}
\end{equation}

The terms in this functional represent the local density approximations for the so-called \textit{ideal gas} contribution and the contribution due to repulsive interactions. The attractive contribution, which contains implicitly the activity, contributes in (\ref{eq:rho_open}) the necessary terms for our model to be of AMB+ form \cite{Tjhung2018}. 

\begin{figure}[t!]
\centering
\includegraphics[width=0.45\textwidth]{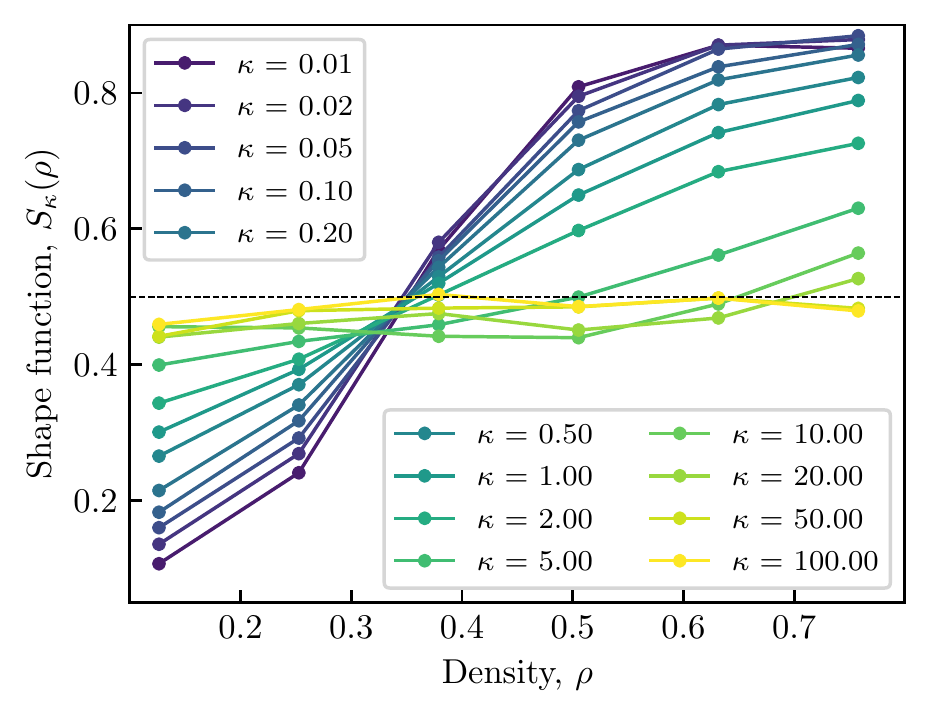}		
\caption{\textit{Measuring $S_\kappa(\rho)$ numerically for non-dimensional switching rates $\kappa \in [10^{-2},10^2]$}. Shape function $S_\kappa(\rho)$ measured from simulations of the microscopic model as the fraction of \textit{on} agents in circular regions of radius $\sigma_c$ with number of agents per unit area $\rho$.}
\label{fig:srho}
\end{figure}

To show this, we first write Eq.\,(\ref{eq:rho_open}) in closed form. The density of \textit{on} particles is related to the density of all particles via $\rho_1(\mathbf{r}) = \mathbb{P}(\epsilon = 1|\rho(\mathbf{r})=\rho_b) \times \rho(\mathbf{r})$, where this conditional probability can be seen as the fraction of particles in a region of bulk density, $\rho\equiv\rho_b$ with internal variable $\epsilon = 1$.   We argue that this conditional probability is a function of the switching rate and the local total density. We write $\rho_1(\rv) = S_{k}(\rho) \rho(\rv)$; we measure $S_{k}(\rho)$ numerically in the simulations of our microscopic model for a wide range of switching rates as shown in Fig.\,\ref{fig:srho}. Here, we made a local density approximation and implicitly assume that the shape function $S_k(\rho)$ does not depend on the gradient of the density field \cite{Barrat2003}.

\textit{Fast Switching Limit pertains to Model B ---} 
If switching happens much faster than diffusion, $k \gg D/\sigma^2$, then we argue that there should be no correlation between the particles local density and their state; we conclude that for large switching rates, $S_{k}(\rho) \equiv \frac{1}{2}$. In this case, we absorb the contribution of the attractive interactions to the probability current in a re-defined free-energy leading to a density equation of \textit{Model B} form (see \cite{Note1} for a full derivation). We conclude that the phase separation in this limit is driven by an effective attraction (with reduced strength) between any pair of agents leading to full phase separation as predicted by \textit{Model B} \cite{Hohenberg1977}.
	
\textit{Fast (but finite) switching leads to Active Model B+ form ---} 
Here, we work perturbatively around the fast switching limit. When considering large but finite values of $k$, we perturb the shape function to linear order and write $S_{k}(\rho) = 1/2 + A_{{k}}(\rho-\rho_c)$. In doing so, we are implicitly modelling a small amount of demixing due to the finite switching rates. After substituting this linear perturbation in the convolution in Eq.\,(\ref{eq:rho_open}) and taking a gradient expansion of the non-local terms, we can re-write the contribution of the attractive interactions to the current up to $\mathcal{O}(\nabla^4\rho^3)$. The coefficients of each of the TRS breaking terms are proportional to $\mu_{k} = \frac{A_{k}}{8}\big(2\rho_cA_{k}-1\big) \int d\rv\:U_1(r)r^2$ \cite{Note1}. Finally, we use the fact that adding a term of the form $\alpha\rho|\nabla\rho|^2$ to the functional $\mathcal{F}[\rho(\rv)]$ generates terms proportional to $\alpha\nabla\big(|\nabla\rho|^2\big) - 2\alpha(\nabla\rho)\nabla^2\rho - 2\alpha\rho\nabla^3\rho$ in the current. Choosing $\alpha=-3\mu_{k}/2$ and again re-defining $\mathcal{F}[\rho]$, we write the density equation in the form 
\begin{equation}
\begin{aligned}
\partial_t \rho(\rv, t)= \nabla\cdot \bigg[\rho(\rv)\bigg(\nabla\bigg(\frac{\delta\mathcal{F}[\rho(\rv)]}{\delta\rho(\rv)}-\frac{5\mu_{k}}{2}|\nabla\rho(\rv)|^2&\bigg)\\ 
											+\mu_{k}\big(\nabla^2\rho(\rv)\big)\nabla\rho(\rv)&\bigg)\bigg]
\end{aligned}
\end{equation}
Finally, we conclude that our model belongs to the AMB+ class, with constants $\lambda = -5\mu_{k}/2 < 0$ and $\zeta = -\mu_{k} < 0$ in the notation of Ref.\,\cite{Tjhung2018}. 
	
\begin{figure}[t!]
\centering
\includegraphics[width=0.45\textwidth]{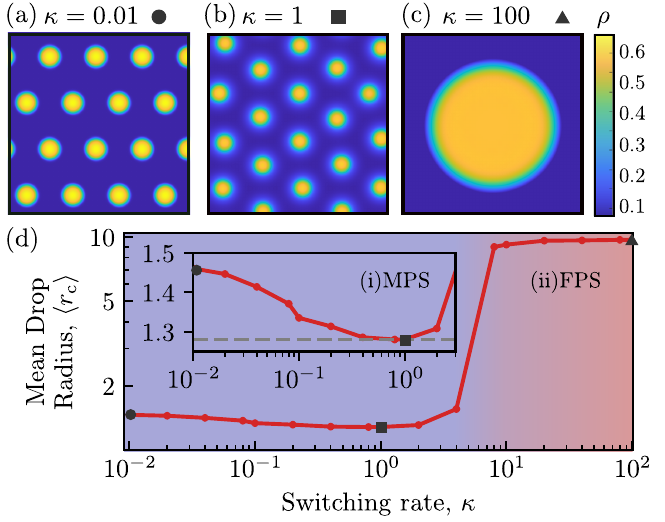}
\caption{\textit{Numerical Analysis of Kinetic Equations}. Numerical solutions of Eqs. (\ref{eq:kineticequations}) for $\bar{\rho} = 0.16$ with $\varepsilon\gg k_BT$  \cite{Note1}. Steady-state solutions show microphase separation (MPS) for (a) $\kappa=0.01$ and (b) $\kappa=1$ but full phase separation (FPS) for (c) $\kappa = 100$. (d) Mean droplet radius $\langle r_c \rangle$ is non-monotonic in the switching rate $\kappa$, in agreement with Fig.\,\ref{fig:emergentstructure}(c).}
\label{fig:ostwald}
\end{figure}

\textit{Active switching drives microphase separation ---} 
We expect to observe the emergence of microphase separation for a range of switching rates. We confirm this by numerically solving the R-DDFT equations (\ref{eq:kineticequations}) \cite{Note1, Roth2010, Hermann2019}. Specifically, we fix the total density of agents $\bar{\rho}$ and size of the solution domain and vary the switching rate $\kappa$. We set $\epsilon\gg k_bT$ as to ensure phase separation from a nearly-homogenous initial condition. For moderate switching rates, the system's steady state supports the coexistence of droplets driven by a reversal of Ostwald ripening \cite{Note1} (Fig.\,\ref{fig:ostwald}). Interestingly, droplet sizes are non-monotonically controlled by the switching rate, $\kappa$. At higher switching rates, we observe full phase separation characterized by a single drop in the solution domain. 

Note that the suppression of Ostwald ripening was first discussed in the context of active emulsions \cite{Zwicker2014,Zwicker2015,Wurtz2018}. We confirm these phenomenological results through the proper coarse-graining of a minimal microscopic model. As argued above, at a macroscopic level, our system can be seen as a binary fluid driven away from equilibrium by an active switching between the two components.

\textit{Microphase separation and Active Model B+} ---
Finally, we connect our two main results: our derivation of the AMB+ density equation used a perturbation of the fast-switching limit while the presence of microphase separation was shown for moderate switching rates. In particular, our linear approximation of the shape function is valid for any $k$, provided that $|\rho-\rho_c|$ is small enough. This is sufficient to conclude on the classification of AMB+ for all switching rates $k>0$ \cite{Note1}.
	
To identify the conditions for microphase separation, we need to go beyond this linear perturbation. To do so, we make an ansatz for the functional form of $S_k(\rho)$ motivated by our computational results (Fig.\,\ref{fig:srho}) that we argue is valid for all $k$. Using this ansatz, we evaluate the coefficients of the TRS breaking terms and compare them to  \cite{Tjhung2018} in which microphase separation in the (deterministic) AMB+ equation was first studied. We find our results to be consistent for all switching rates \cite{Tjhung2018, Note1}.

Namely, for infinitely fast switching, one recovers an effective equilibrium field theory of Model B type with reduced attractive interactions. For fast (but finite) switching, shallow gradients in the shape function imply negative but small coefficients of the TRS breaking terms leading to FPS. For moderate switching rates, the gradient steepens leading to large and negative coefficients leading to ROR and MPS. We argue that $k$ controls how deep in the MPS region the system is and that the non-monotonic dependence of the cluster size stems from the non-monotonic behavior of $\mu_k$ as we decrease $k$ \cite{Note1}.

\textit{Discussion ---} Using a bottom-up approach, we introduce a minimally active microscopic model inspired by the pili-mediated interactions of \textit{Neisseria Meningitidis}. Through a rigorous coarse-graining procedure, we show that its density equation is of \textit{Active Model B+} form. While, until now, only motile active matter systems were shown to produce the necessary time-reversal symmetry breaking terms, we formally link this field theory to a non-motile active model \cite{Tjhung2018}.  Further, we reveal in our model the existence of microphase separation, controlled by the switching rate $k$. In the context of bacteria dynamics, intermittent attractive interactions mediated by pili dynamics lead to a mechanism controlling bacterial clustering and regulating typical cluster sizes. More generally, our work lays the foundations to study non-equilibrium phase separation in field theories. In a field dominated by dry motile active matter models, our {\it truly} scalar microscopic model offers a new path to model wet active matter. While our study focuses on bacterial clustering, we believe our model has much wider applications and can for instance be used to model the dynamics of eukaryotic spheroids, in which our fluctuating forces would capture intercellular tension fluctuations \cite{Oriola2021,Kim2021}.


%

\end{document}



\title{Supplemental Information for ``Intermittent Attractive Interactions Lead to Microphase Separation in Non-motile Active Matter"}

\author{Henry Alston}
\affiliation{Department of Mathematics, Imperial College London, 180 Queen's Gate, London SW7 2BZ, United Kingdom}
\author{Andrew O. Parry}
\affiliation{Department of Mathematics, Imperial College London, 180 Queen's Gate, London SW7 2BZ, United Kingdom}
\author{Rapha\"el Voituriez}
\affiliation{Laboratoire de Physique Th\'eorique de la Mati\`ere Condens\'ee, UMR 7600 CNRS/UPMC, 4 Place Jussieu, 75255 Paris Cedex, France}
\affiliation{Laboratoire Jean Perrin, UMR 8237 CNRS/UPMC, 4 Place Jussieu, 75255 Paris Cedex, France}
\author{Thibault Bertrand}
\email{t.bertrand@imperial.ac.uk}
\affiliation{Department of Mathematics, Imperial College London, 180 Queen's Gate, London SW7 2BZ, United Kingdom}
	
\date{\today}
	
\maketitle
\hrule
\tableofcontents
\vspace{1em}
\hrule	


\vspace{1cm}

\section{Numerical analysis of the microscopic model}

\subsection{Non-dimensionalization of the microscopic model}

We consider a system of $N$ unit mass particles of size $\sigma$ characterised by their position $\rv_i$ and the value of an internal variable $\epsilon_i\in\{0,1\}$ defining their interactions. The dynamics in our microscopic model of interacting particles are governed by the following overdamped Langevin equation:
\begin{equation}
\dot{\rv}_i = -\frac{1}{\gamma}\sum_{j\ne i} \nabla_{\rv_i} U(r_{ij}, \epsilon_i\epsilon_j) + \sqrt{2D}\bm{\eta}_i,
\label{seq:langevin}
\end{equation}
where $\gamma$ is a friction coefficient, $D$ is the bare-diffusion coefficient which sets the temperature in the system (through Einstein's relation), $U$ is the total pair interaction potential defined as the superposition of a purely repulsive contribution $U_0$ and a purely attractive contribution $U_1$:
\begin{equation}
U(r_{ij}, \epsilon_i\epsilon_j)=U_0(r_{ij})+ \epsilon_i\epsilon_j U_1(r_{ij})
\label{seq:fullpotential}
\end{equation}
and $\bm{\eta}_i$ as a zero mean, unit variance Gaussian white noise term such that 
\begin{align}
\left\langle \eta^{\alpha}_i(t) \right\rangle  &= 0  \\
\left\langle \eta^{\alpha}_i(t) \eta^{\beta}_j(t') \right\rangle  &= \delta_{ij}\delta_{\alpha\beta}\delta(t-t')  
\end{align}
with $\alpha,\beta \in \{x,y\}$.

The pair interaction potential is defined such that: (1) any two particles interact through steric repulsion when their center-to-center distance is such that $|\rv_i-\rv_j|=r_{ij}<\sigma_* = 2^{1/6}\sigma$ (independently of the value of $\epsilon_i$ and $\epsilon_j$) and (2) if the internal variables of both agents are such that $\epsilon_i=\epsilon_j=1$, these particles are {additionally} subjected to an attractive force with longer range $\sigma_c = 2\sigma$. Following the classical WCA potential decomposition for the Lennard Jones potential \cite{Weeks1971, Barrat2003}, we define the two contributions to the pair interaction potential as follows: 
\begin{align}
U_0(r_{ij}) &= \begin{cases}
4\varepsilon\left[\left(\sigma/r_{ij}\right)^{12} - \left(\sigma/r_{ij}\right)^6 \right] + \varepsilon, & r_{ij}<\sigma_*\\
0, & r_{ij}\ge\sigma_*
\end{cases}
\\
U_1(r_{ij}) &= \begin{cases}
- \varepsilon, & r_{ij}<\sigma_*\\
4\varepsilon\left[\left(\sigma/r_{ij}\right)^{12} - \left(\sigma/r_{ij}\right)^6 \right], & \sigma_* \le r_{ij}\le \sigma_c.
\end{cases}
\end{align}
 
Finally, activity is introduced by allowing the particles to stochastically jump between the {\it on} and {\it off} states with constant rates which leads to intermittent attractive forces. In practice, the internal variables $\{\epsilon_i\}_{i\in[1,N]}$ follow independent two-state Markov processes (or telegraph process) with switching rates $k_{\rm on}$ and $k_{\rm off}$, leading to
\begin{equation}
\epsilon_i = 0 \xrightleftharpoons[k_{\rm off}]{k_{\rm on}} \epsilon_i = 1
\end{equation}
and we formally write 
\begin{subequations}
\begin{align}
\p_t P(\epsilon_i = 0) &= - k_{\rm on} P(\epsilon_i = 0) + k_ {\rm off} P(\epsilon_i = 1), \\
\p_t P(\epsilon_i = 1) &=   k_ {\rm on} P(\epsilon_i = 0) - k_{\rm off} P(\epsilon_i = 1).
\end{align}
\label{eq:switchrates}	
\end{subequations}
These dynamics lead to exponentially distributed waiting times between switching events with respective timescales $k_{\rm on}^{-1}$ and $k_{\rm off}^{-1}$. 

We non-dimensionalize our equations using the following scheme:
\begin{equation}
\rv^*=\frac{\rv}{\sigma}, \quad U^*=\frac{U}{\varepsilon} \quad\text{and}\quad t^* = \frac{Dt}{\sigma^2} 
\end{equation}
where we use $\sigma$ the nominal particle diameter and $\varepsilon$ the depth of the attractive interaction as length and energy scales, respectively. These are natural choices stemming from the Lennard-Jones potential. We define our timescale as $\tau = \sigma^2/D$, where again $D$ is the bare-diffusion coefficient which sets the temperature in the system through Einstein's relation. Provided these length-, time- and energy scales, we can non-dimensionalize our equation of motion (\ref{seq:langevin}) to obtain
\begin{equation}
\dot{\rv}^*_i = -\Gamma\sum_{j\ne i} \nabla_{\rv^*_i} U^*(r_{ij}, \epsilon_i\epsilon_j) + \sqrt{2}\bm{\eta}^*_i.
\label{seq:ndlangevin}
\end{equation}
Note that the non-dimensionalization of the noise term in (\ref{seq:langevin}) requires special care. As our equation is dimensionally homogeneous, we have $\left[ \sqrt{2D} \bm{\eta}_i\right] = L T^{-1}$, which implies by dimensional analysis that $\left[\bm{\eta}_i\right] = T^{1/2}$. We can indeed confirm that this is correct by studying the noise correlation function:
\begin{equation}
\left\langle \eta^{\alpha}_i(t) \eta^{\beta}_j(t') \right\rangle = \delta_{ij}\delta_{\alpha\beta}\delta(t-t')   \quad\implies \int d\tau \left\langle \eta^{\alpha}_i(t) \eta^{\alpha}_i(t+\tau) \right\rangle = 1.
\end{equation} 
It follows that the noise correlator has the units of the inverse of a time and thus the noise has units $T^{-1/2}$. Hence, we non-dimensionalize it as
\begin{equation}
\bm\eta^* = \frac{\sigma}{\sqrt{D}}\bm\eta.
\end{equation}

Our non-dimensionalisation  leaves us with a single free dimensionless parameter $\Gamma = \varepsilon/\gamma D = \varepsilon/ k_B T$, where the second equality holds by Einstein's relation. As we are interested in systems displaying phase separation, we generically set $\Gamma \gg 1$ in our simulations. Indeed at low values of $\Gamma$, attraction is weak compared to the thermal fluctuations, preventing sustained phase separation.

Similarly, the pair potential of the particles non-dimensionalizes to 
\begin{align}
U^*_0(r^*_{ij}) &= \begin{cases}
4\left[\left(r^*_{ij}\right)^{-12} - \left(r^*_{ij}\right)^{-6} \right] + 1, & r^*_{ij}<2^{1/6}\\
0, & r^*_{ij}\ge2^{1/6}
\end{cases}
\\
U^*_1(r^*_{ij}) &= \begin{cases}
- 1, & r^*_{ij}<2^{1/6}\\
4\left[\left(r^*_{ij}\right)^{-12} - \left(r^*_{ij}\right)^{-6} \right] , & 2^{1/6} \le r^*_{ij}\le 2.
\end{cases}
\end{align}
Throughout the main text and supplementary information, we drop the asterisk notation used above to indicate non-dimensional variables to make for easier reading. Numerical results are stated in non-dimensional variables whereas the analytic derivations work in dimensional variables. Finally, we non-dimensionalize the switching rates using the same timescale as above, which leads us to define the non-dimensional switching rates $\kappa_{\rm on/off}=k_{\rm on/off}\sigma^2/D$.

\subsection{Computational details}
	
We solve the non-dimensionalized Langevin equation (\ref{seq:ndlangevin}) using a stochastic Runge-Kutta scheme \cite{Branka1999} with a maximum time step $dt=1\times10^{-5}\tau$. We use periodic boundary conditions in a square box. The size of the box is set by $\phi$ the total volume fraction of agents fixing the side length of the solution domain to $L = \sqrt{N\pi\sigma_*^2/4\phi}$, with the number of particles $N=2000$ unless stated otherwise. Unless stated otherwise, our simulations are performed at an average volume fraction $\phi=0.3$ to allow for the coexistence of regions of high and low density. To increase the efficiency of our simulations, we use a hybrid neighbor-list method to compute particle interactions based on combined cell-linked lists and Verlet neighbor lists. 

All our simulations were ran for at least $500\tau$ which we found to be long enough for the system to reach a non-equilibrium steady state. We checked this by measuring the radius of gyration, $R_{\rm{gyr}}$ (see below). As the system reaches steady-state the value of the radius of gyration plateaus. For low switching rates, we run the simulations for at least $(10\kappa^{-1})\tau$ which we argue is long enough such that the position and state of each particle at the end of the simulations had lost its dependence on their initial values. 
	
Unless stated otherwise, our microscopic numerical simulations are initialized by seeding the particles such that they form a single cluster. Over time, agents may diffuse away from the initial drop depending on the conditions and the existence of stable dense and dilute phases. Using these initial conditions allows us to vastly reduce the time the system requires to reach stationarity.
	
\subsection{Measuring structural characteristics}\label{ssec:structquant}
	
The \textbf{radius of gyration}, $R_{\rm{gyr}}$, is defined as 
\begin{equation}
R_{\rm{gyr}}(t) = \sqrt{\frac{1}{N} \sum_{i} |\rv_i(t)- \rv_{\rm cm}(t)|},
\end{equation}
where $\rv_{\rm cm}(t) = N^{-1} \sum_{i=1}^N \rv_i(t)$ is the position of the center of mass of the particles at time $t$. We use the radius of gyration to determine whether the system exhibits full liquid-gas phase separation. In the case where all particles are clustered in a single drop, we expect this measure to be low as all particles are relatively close to the center of mass of the cluster. Conversely, the radius of gyration increases in systems where particles are uniformly distributed across the simulation domain. Note that we use in all our simulations periodic boundary conditions; in simulations leading to a dispersed phase, we thus expect the radius of gyration to be monotonically increasing over time as the original cluster expands but to remain bounded, i.e. it will converge to a finite value set by the size of the simulation box in the limit $t\rightarrow\infty$.

To quantify the amount of demixing between {\it on} and {\it off} states particles, we use a {\bf demixing index}, ${I_{\rm{demix}}}$ defined as the fraction of Voronoi neighbors with the same state, i.e. 
\begin{equation}
I_{\rm demix} = \left\langle N_i^{s} / N_{i}^t\right\rangle_i
\end{equation}
where $N^s_i$ and $N^t_i$ are the number of same type and total number of neighbouring particles, respectively, and the average runs over all particles. For a fully demixed system, we would expect an index  approaching $1$, while a fully mixed system with equal switching rates leads to $I_{\rm{demix}}=0.5$.

Finally, we measure the \textbf{maximal cluster size}, $s_{\rm{max}}$. At any given time, we proceed to a cluster size analysis; clusters are defined as sets of particles within interaction range $\sigma_c$ of each other.

These measures are used to distinguish the different emergent structures observed in our system as we vary microscopic parameters. For instance, we identify a microphase separated system as one with high radius of gyration (which implies no \textit{full} phase separation) along with moderate maximal cluster size indicating some level of liquid-gas phase separation present in the system.  

\subsection{Measuring $S_\kappa(\rho)$ numerically from simulations}
	
To measure the shape function, $S_\kappa(\rho)$, we take a snapshot of the system at a non-equilibrium steady state for a chosen value of the non-dimensional switching rate, $\kappa$. We then take circular samples of the system with radius $2\sigma_*=\sigma_c$ to replicate the range of interaction for any particle, being careful with boundary conditions. For each sample, we count the number of agents whose center is in the region and the fraction of those which are in an \textit{on} state. After repeating this for many samples from different snapshots of the system, we calculate the fraction of agents which are in an \textit{on} state for each band of the local density, $\rho$.
	
\section{Deriving R-DDFT kinetic equations from many-body Smoluchowski equation}
\label{ssec:Smoldev}

From the equations for our microscopic model, we write down a many-body Smoluchowski equation for the $N$-agent distribution function, which we denote by $\psi_N$. We set $\gamma=1$ and write
\begin{equation}\begin{aligned}\label{seq:mbsmol}
\partial_t\psi_N(\{\rv_1, \dots, \rv_N, \epsilon_1, \dots, \epsilon_N\},t) = \sum_{n=1}^N \:\bigg[&\nabla_{\rv_n}\cdot\bigg[\sum_m\nabla_{\rv_n}U(r_{nm}, \epsilon_n\epsilon_m)+D\nabla_{\rv_n}\bigg] \psi_N+k_1(\epsilon_n)S_n\psi_N-k_2(\epsilon_n)\psi_N\bigg]
\end{aligned}
\end{equation}
where $r_{nm} = |\rv_n-\rv_m|$ and we have defined
\begin{equation}
k_1(\epsilon) = \epsilon k_{\rm{on}} + (1-\epsilon)k_{\rm{off}} \quad and \quad k_2(\epsilon) = k_1(1-\epsilon) 
\end{equation}
and the switch operator $S_i$ is defined as follows 
\begin{equation}
S_i\psi_N= \psi_N(\rv_1, \dots,  \rv_N, \epsilon_1, \dots, 1-\epsilon_i, \dots, \epsilon_N).
\end{equation}
Note that by definition $S_i^2\psi_N = \psi_N$.  We define the single-agent distribution function, $\psi_1$, as
\begin{equation}
\psi_1(\rv, \epsilon, t) = \sum_{\epsilon_2=0}^1\dots\sum_{\epsilon_N=0}^1 \int d\rv_2\dots\int d\rv_N N\psi_N,
\end{equation}
where we have dropped the subscript on the tagged particle. The factor of $N$ is present as we can choose any one of the $N$ particles. From this definition, we can derive the evolution equation for the single-agent distribution function from (\ref{seq:mbsmol}) by integrating over the remaining position and state variables:
\begin{equation}
\begin{aligned}\label{seq:s11}
\partial_t\psi_1(\rv,\epsilon, t) &= -\nabla_{\rv}\cdot\mathbf{F}(\rv,\epsilon,t)+D\Delta_{\rv}\psi_1+k_1(\epsilon)S_1\psi_1-k_2(\epsilon)\psi_1\\
							&= \nabla_{\rv} \cdot \big[D\nabla_{\rv}\psi_1 -\mathbf{F}(\rv,\epsilon,t) \big] +k_1(\epsilon)S_1\psi_1-k_2(\epsilon)\psi_1
\end{aligned}
\end{equation}
where we have introduced the mean force on the tagged particle, $\mathbf{F}$, as
\begin{equation}
\begin{aligned}    
\mathbf{F}(\rv,\epsilon,t)&= \sum_{\epsilon_2=0}^1\dots\sum_{\epsilon_N=0}^1 \int d\rv_2 ~\dots\int d\rv_N \bigg[-\sum_{j=2}^N\nabla_{\rv}U(|\rv-\rv_j|, \epsilon\epsilon_j)\bigg]N\psi_N\\
						&=\sum_{\epsilon'=0}^1\int d\rv' \big[-\nabla_{\rv}U(r, \epsilon\epsilon')\big]\psi_2(\rv,\rv',\epsilon,\epsilon', t)
\end{aligned}
\end{equation}
for $r=|\rv-\rv'|$ and the two agent distribution function, $\psi_2$. 

The mean force term here can be written as the sum of two components, accounting for repulsive and attractive interactions: 
\begin{equation}\begin{aligned}\label{seq:meanforce}
\textbf{F}(\rv,\epsilon, t)= \sum_{\epsilon'=0}^1\int d\rv' \big[-\nabla_{\rv}U_0(r)\big]\psi_2(\rv, \rv', \epsilon, \epsilon', t) + \int d\rv' \big[-\nabla_{\rv}U_1(r)\big]\psi_2(\rv, \rv', \epsilon, 1, t).
\end{aligned}
\end{equation}
We consider the mean force for each value of the internal variable, $\epsilon$, independently: 
\begin{align}
\textbf{F}(\rv,0, t)&=\sum_{\epsilon'=0}^1\int d\rv' \big[-\nabla_{\rv}U_0(r)\big]\psi_2(\rv,\rv',0,\epsilon', t)  = \int d\rv' \big[-\nabla_{\rv}U_0(r)\big]\tilde{\psi}_2(\rv,\rv',0, t)\\
\textbf{F}(\rv,1, t)&=\int d\rv' \big[-\nabla_{\rv}U_0(r)\big]\tilde{\psi}_2(\rv,\rv',1, t) + \int d\rv' \big[-\nabla_{\rv}U_1(r)\big]\psi_2(\rv, \rv', 1, 1, t) 
\end{align}
where we have defined the following notation $\tilde{\psi}_2(\rv, \rv', \epsilon,t) = \sum_{\epsilon'=0}^1 \psi_2(\rv, \rv', \epsilon, \epsilon', t)$. To simplify the form of this mean force, we make use of two approximations commonly introduced in \textit{Dynamic Density Functional Theory} (DDFT) \cite{Archer2004, Lowen1994, Barrat2003, TeVrugt2020a}. We first simplify the contribution due to repulsive interactions in the equation for the density of \textit{off} ($\epsilon=0$) agents and \textit{on} ($\epsilon=1$) agents, respectively. Finally, we detail the mean field approximation used to capture the attractive contributions to the density equation for \textit{on} agents. Note that we define $\rho_i(\rv, t) = \psi_1(\rv, i, t)$ for $i\in\{0,1\}$ as the density of \textit{off} and \textit{on} agents, respectively.
	
\subsection{Mean force for \textit{off} agents} 

We now replace the time-dependent, two agent density with the equivalent density at equilibrium, $\tilde{\psi}_2(\rv, \rv', 0).$ This assumption is valid at high density where the steric repulsion highly dictates the particle structure in the system \cite{Barrat2003}. This allows us to write the mean force as the gradient of the single-particle direct correlation function for an inhomogeneous fluid, $c_{1}(\rv)$, as 
\begin{equation}
\int d\rv' \big[-\nabla_{\rv}U_0(r)\big]\tilde{\psi}_2(\rv, \rv', 0)= k_B T \rho_0(\rv) \nabla_{\rv}\big(c_{1}(\rv)\big).
\end{equation}
where $\rho_0(\rv)$ is the density of {\it off} particles at position $\rv$. 
			
Following \textit{Density Functional Theory} (DFT), we connect the single particle direct correlation functional to the excess free-energy for the repulsive interactions: $\mathcal{F}^{\rm{ex}}_{\rm{rep}}[\rho(\rv)]$. The result holds exactly at equilibrium; we make the assumption that it carries over to the non-equilibrium case with the same excess free energy functional. This allows us to write the contribution due to repulsive interactions as 
\begin{equation}
k_BT\rho_0(\rv)\nabla_{\rv}\big(c_{1}(\rv)\big) = -\rho_0(\rv)\nabla\bigg(\frac{\delta \mathcal{F}^{\rm{ex}}_{\rm{rep}}[\rho(\rv)]}{\delta \rho(\rv)}\bigg).
\end{equation}  
For the more familiar form of the DDFT equation as derived in \cite{Archer2004}, we can consider a local density approximation for the free energy:
\begin{equation}
\mathcal{F}^{\rm{ex}}_{\rm{rep}}[\rho(\rv)] = \int d\rv\:f_{\rm{rep}}\big(\rho(\rv)\big), \quad \frac{\delta \mathcal{F}^{\rm{ex}}_{\rm{rep}}[\rho(\rv)]}{\delta \rho(\rv)} = f'_{\rm{rep}}\big(\rho(\rv)\big)
\end{equation}
and define a chemical potential $\mu_{\rm{rep}}(\rho(\rv))=f'_{\rm{rep}}\big(\rho(\rv)\big)$. In this notation, the repulsive component of the mean force is written as $-\rho_0(\rv)\nabla_{\rv}\mu_{\rm{rep}}(\rho(\rv))$.
			
\subsection{Mean force for \textit{on} agents}

The repulsive interactions contribution to the mean force for \textit{on} agents is of the same form as that derived for the \textit{off} agents: $-\rho_1(\rv)\nabla_{\rv}\mu_{\rm{rep}}(\rho(\rv))$, where $\rho_1(\rv)$ is now the density of {\it on} particles at $\rv$. Note that the chemical potential is a function of the total density of agents in each case. This makes sense as any pair of agents interact through steric repulsion. 
			
Focusing now on the attractive contribution to the mean force, we again assume that the instantaneous two agent distribution function is replaced by its equilibrium counterpart. To re-write this attractive contribution, we use the following approximation: the attractive interaction potential, $U_1$, is weak compared to the divergent repulsive potential, $U_0$. It is well understood that a mean field approximation for the two agent distribution function is accurate for a wide range of temperatures and density for weak (or soft) potentials \cite{Barrat2003}. This mean field approximation implicitly assumes that the positions of pairs of agents interacting through this potential are uncorrelated across the system. 

We thus write the two agent distribution function as
\begin{equation}
\psi_2(\rv, \rv', 1, 1) \approx \frac{1}{2} \rho_1(\rv)\rho_1(\rv').
\end{equation}
We substitute it in to the attractive component of (\ref{seq:meanforce}) and write			
\begin{equation}
\int d\rv' \big[-\nabla_{\rv}U_1(r)\big]\psi_2(\rv, \rv', 1, 1) = -\frac{1}{2}\rho_1(\rv)\nabla_{\rv}\big(U_1\star \rho_1\big)
\end{equation}
where the $\star$ here represents a convolution integral. We can define a chemical potential as we did for the repulsive interaction in the form $\mu_{\rm{atr}}(\rho_1(\rv)) = \frac{1}{2}\big(U_1\star \rho_1\big)$ and write the attractive contribution as $-\rho_1(\rv)\nabla_{\rv}\mu_{\rm{atr}}(\rho_1(\rv))$. Our result is of the same form as in \cite{Dean1996} where only soft interaction potentials were considered.
				
\subsection{Kinetic equations}

We can now write the kinetic equations for our model. From (\ref{seq:s11}), we conclude that the time derivatives of the {\it on} and {\it off} particles densities can be written as 
\begin{subequations}
\begin{align}
\partial_t \rho_0(\rv, t)  &= \nabla_{\rv}\cdot \Big[D\nabla_{\rv}\rho_0(\rv) + \rho_0\nabla_{\rv}\mu_{\rm{rep}}(\rho(\rv))\Big] + k \rho_1(\rv) - k\rho_0(\rv) \label{seq:rddft1}\\
\partial_t \rho_1(\rv, t)  &= \nabla_{\rv}\cdot \Big[D\nabla_{\rv}\rho_1(\rv) + \rho_1\nabla_{\rv}\big(\mu_{\rm{rep}}(\rho(\rv))+\mu_{\rm{atr}}(\rho_1(\rv))\big)\Big] + k \rho_0(\rv) - k\rho_1(\rv) \label{seq:rddft2}
\end{align}
\label{seq:rddft}	
\end{subequations}
justifying the R-DDFT approach used in the main text \cite{TeVrugt2020a, TeVrugt2020b,MonchoJorda2020, Zakine2018}.

\section{Fast switching pertains to Model B}

We show how to derive a density equation of Model B form from our R-DDFT equations. We take the sum of the two equations (\ref{seq:rddft1}-\ref{seq:rddft2}) to write an evolution equation for the total density of agents. Following the literature for equilibrium field theories \cite{Wittkowski2014, Tjhung2018, Bray2002}, we write the passive contributions, the diffusive and repulsive terms, in terms of a free-energy-like functional. We also write $\mu_{\rm{atr}}(\rho(\rv))$ in its full form. The resulting equation is
\begin{equation}\label{deqn}
\partial_t  \rho(\rv, t) = \nabla\cdot\bigg[ \rho(\rv)\nabla\bigg(\frac{\delta\mathcal{F}[ \rho(\rv)]}{\delta  \rho(\rv)}\bigg) + \frac{1}{2}\rho_1(\rv)\nabla\big(U_1\star\rho_1\big)\bigg].
\end{equation}
As in the main text, we use the approximation of a \textit{shape} function to write a closed equation for the density of agents: $\rho_1(\rv) = S_{k}(\rho(\rv))\rho(\rv)$. In the fast switching limit, we argue that this function is approximately $\frac{1}{2}$ (for equal switching rates). We confirm this by measuring the shape function numerically from our simulations of the microscopic model. 
	
We can then write a closed equation for the density of the form
\begin{equation}\label{modB1}
\partial_t  \rho(\rv, t) = \nabla\cdot\bigg[ \rho(\rv)\nabla\bigg(\frac{\delta\mathcal{F}[ \rho(\rv)]}{\delta  \rho(\rv)}\bigg) + \frac{1}{8}\rho(\rv)\nabla\big(U_1\star\rho\big)\bigg].
\end{equation} 
The contribution due to attractive interactions can now be included in a re-defined free energy functional. Indeed, we define the new functional $\mathcal{F}_B[\rho(\rv)]$ as 
\begin{equation}
\mathcal{F}_B[\rho(\rv)] = \mathcal{F}[\rho(\rv)] + \frac{1}{8}\int d\rv \big[\rho(\rv)\big(U_1\star\rho)\big]
\end{equation}
such that (\ref{modB1}) can be written as 
\begin{equation}\label{eqn:modelB}
\partial_t  \rho(\rv, t) = \nabla\cdot\bigg[ \rho(\rv)\nabla\bigg(\frac{\delta\mathcal{F}_B[ \rho(\rv)]}{\delta  \rho(\rv)}\bigg)\bigg].
\end{equation} 
This is a density equation of \textit{Model B} form. We can recover the classical \textit{Model B} form by rescaling the density and defining a scalar order parameter $\phi(\rv, t) = (\rho(\rv, t) - \rho_c)/\rho_c$, where $\rho_c$ is the critical density for the system. A similar derivation is found in \cite{Tjhung2018}.

\section{Fast (but finite) switching leads to Active Model B+ form}

\subsection{Derivation of Active Model B+ density equation for linear approximation of $S_k(\rho)$}
\label{ssec:ambpderiv}

{In this section, we proceed to the derivation of \textit{Active Model B+} equation from R-DDFT equations.} The density equation for an \textit{Active Model B+} system is written in the form
\begin{equation}
\partial_t \rho(\rv, t) = \nabla\cdot\bigg[\rho(\rv)\bigg(\nabla\bigg(\frac{\delta\mathcal{F}(\rho(\rv))}{\delta\rho(\rv)}+\lambda|\nabla\rho(\rv)|^2\bigg) - \zeta \nabla\rho(\rv)\nabla^2\rho(\rv)\bigg)\bigg].
\end{equation}
where $\lambda$ and $\zeta$ are constants and $\mathcal{F}[\rho(\rv)]$ contains all the passive terms in the current \cite{Tjhung2018}. 
	
From (\ref{deqn}) and substituting our expression for $\rho_1(\rv)$, we write
\begin{equation}\label{eq:closedDensity}
\partial_t \rho(\rv, t)= \nabla\cdot \Bigg[\rho(\rv)\nabla\bigg(\frac{\delta\mathcal{F}(\rho(\rv))}{\delta\rho(\rv)}\bigg)+\frac{1}{2}{\rho(\rv)S_k(\rho(\rv))}\nabla\bigg(\int d\rv'\:U_1(|\rv'-\rv|)\rho(\rv')S_k(\rho(\rv'))\bigg)\Bigg].
\end{equation}
We work perturbatively around the fast switching limit and expand the shape function up to linear order in $\rho$:
\begin{equation}
S_{k}(\rho(\rv)) = \frac{1}{2} + A_{k}(\rho(\rv) - \rho_c) = \frac{1}{2}-A_{k}\rho_c + A_{k}\rho(\rv).
\end{equation}
For simplicity, we write $B_{k}=\frac{1}{2}-A_{k}\rho_c$
and use a gradient expansion for the non-local terms in our system (terms evaluated at $\rv'$ and not $\rv$):
\begin{equation}
\rho(\rv') = \rho(\rv) + \nabla\rho\cdot(\rv'-\rv) + \frac{\nabla^2 \rho}{2}r^2 + \dots
\end{equation}
It follows that 
\begin{equation}
S_k(\rho(\rv')) = B_{k}+A_{k}\rho(\rv') \approx B_{k}+A_{k}\bigg( \rho(\rv) + \nabla\rho(\rv)\cdot(\rv'-\rv) + \frac{\nabla^2 \rho(\rv)}{2}r^2\bigg).
\end{equation}
	
As we are now working with only local terms for the density, we write $\rho=\rho(\rv)$ for simplicity. The convolution term involving the attractive potential, $U_1$, can be written as
\begin{equation}
\int d\rv'\:U_1(|\rv'-\rv|)\bigg( \rho + \nabla\rho\cdot(\rv'-\rv) + \frac{\nabla^2 \rho}{2}r^2\bigg)\bigg(B_{k}+A_{k}\bigg( \rho + \nabla\rho\cdot(\rv'-\rv) + \frac{\nabla^2 \rho}{2}r^2\bigg)\bigg).
\end{equation} 
We define the constants 
\begin{equation}
\alpha_0 = - \int d\rv\: U_1(|\rv|),\quad \alpha_2 = -\int d\rv\: |\rv|^2U_1(|\rv|) 
\end{equation}
such that $\alpha_0, \alpha_2>0$ and evaluate the integral as
\begin{equation}
-\alpha_0\rho(B_{k}+A_{k}\rho)-\alpha_2\bigg(\frac{A_{k}}{2}|\nabla\rho|^2 + A_{k}\rho\nabla^2\rho + \frac{B_{k}}{2}\nabla^2\rho\bigg).
\end{equation}

Returning now to the current in the density equation, we evaluate the total contribution due to attractive interactions as
\begin{equation}
-\frac{\rho(B_{k}+A_{k}\rho)}{2}\nabla \bigg(\alpha_0\rho(B_{k}+A_{k}\rho)+\alpha_2\bigg(\frac{A_{k}}{2}|\nabla\rho|^2 + A_{k}\rho\nabla^2\rho + \frac{B_{k}}{2}\nabla^2\rho\bigg)\bigg).
\end{equation}
To compare this to the \textit{Active Model B+} density equation, we need to first determine which terms can be absorbed in to a re-defined free energy. The term proportional to $\alpha_0$ can be absorbed, so can the term proportional to $B_{k}^2$. Up to order $\mathcal{O}(\nabla^4\rho^3)$, the remaining terms have coefficient proportional to $B_{k}A_{k}\alpha_2/4>0$. We define $\mu_{k} = B_{k}A_{k}\alpha_2/4$ and return to the density equation to write
	
\begin{equation}
\partial_t \rho(\rv, t)= \nabla\cdot \bigg[\rho\bigg(\nabla\bigg(\frac{\delta\mathcal{F}_a[\rho]}{\delta\rho}-\mu_{k}|\nabla\rho|^2\bigg) -2\mu_{k}\nabla\rho\nabla^2\rho-3\mu_{k}\rho\nabla^3\rho \bigg)
\end{equation}
where $\mathcal{F}_a$ is the re-defined free energy after including the terms from the attractive interactions contribution to the chemical potential. Finally, we use the fact that adding a term of the form  $z\rho|\nabla\rho|^2$ to the free energy-like functional generates terms proportional to 
\begin{equation}
+z\nabla\big(|\nabla\rho|^2\big) - 2z(\nabla\rho)\nabla^2\rho - 2z\rho\nabla^3\rho
\end{equation}
in the current of the density equation. We choose $z$ such that $-2z=3\mu_{k}$ and re-define the free energy such that 
\begin{equation}
\partial_t \rho(\rv, t)= \nabla\cdot \bigg[\rho\bigg(\nabla\bigg(\frac{\delta\mathcal{F}_z[\rho]}{\delta\rho}-\frac{5\mu_{k}}{2}|\nabla\rho|^2\bigg) +\mu_{k}\nabla\rho\nabla^2\rho\bigg)\bigg].
\end{equation}
	
We conclude that the system is of \textit{Active Model B+} form with the coefficients of the two time-reversal-symmetry breaking terms given by
\begin{equation}
\lambda = -\frac{5\mu_{k}}{2} < 0 \quad\text{and}\quad \zeta = -\mu_{k} < 0.
\end{equation}
We note that both of these coefficients are related to $A_k$, i.e. to the gradient of the shape function around the critical density $\rho_c$. To get the result in the main text, we recall that $B_k = \frac{1}{2}-A_k\rho_c$, hence $\mu_{k} = \frac{A_{k}}{8}\big(2\rho_cA_{k}-1\big)(-\alpha_2) = \frac{A_{k}}{8}\big(2\rho_cA_{k}-1\big) \int d\rv\:U_1(r)r^2$. 

Here, we worked perturbatively around the fast switching limit. As $k$ decreases, our linear approximation of the shape function only remains valid when $|\rho-\rho_c|$ is sufficiently small. The above results tell us that small fluctuations about the critical density evolve according to an equation of AMB+ form for any $k\in(0,\infty)$, hence we can conclude on AMB+ form for all switching rates.

\subsection{Microphase separation for moderate switching rates}

To conclude on the presence of microphase separation, we need to know more about the specific values of the coefficients for the TRS breaking terms; in particular, microphase separation is only observed for negative and large enough values of the TRS breaking terms coefficients. Going beyond the general linear approximation used above, we here introduce an ansatz motivated by the results of our numerical simulations and write
\begin{equation}
	S_k(\rho) = \frac{1}{2}+\frac{1}{2} \tanh(Y_k(\rho-\rho_c))
\end{equation}  
where the $k$ dependence is contained within the constant $Y_k>0$, which decreases monotonically as $k$ increases. We argue that this approximation is valid for the full range of switching rates, $k$, as opposed to the general linear approximation that we made above. We then expand the $\tanh(\cdot)$ term as 
\begin{align}
	\tanh(Y_k(\rho-\rho_c)) &= Y_k(\rho-\rho_c) - \frac{Y_k^3}{3}(\rho-\rho_c)^3 + \dots \\
	&\label{eq:sums}= \bigg[-Y_k\rho_c + \frac{Y_k^3}{3}\rho_c^3 + \dots\bigg] + \bigg[Y_k - \frac{Y_k^3}{3}(3\rho_c^2) + \dots\bigg]\rho + \mathcal{O}(\rho^2)\\
	&= -\tanh(Y_k\rho_c) +  Y_k(1-\tanh^2(Y_k\rho_c))\rho+\dots
\end{align}
where we have used $Y_k(1-\tanh^2(Y_k\rho_c)) = \partial_{\rho_c}\tanh(Y_k\rho_c)$. Note that the infinite sums in the square brackets only converge if $Y_k<\frac{\pi}{2\rho_c}\approx 4.48$. If we compare our results to Fig.\,3 of the main text, we argue that this condition is satisfied for all the switching rates studied here as the gradient of the shape function at $\rho=\rho_c$ is sufficiently shallow. Note here we are also assuming that $D/\sigma^2 \approx \mathcal{O}(1)$ such that the magnitude of the dimensional switching rate is comparable to its non-dimensionalized rate. Otherwise, there is a scaling factor to consider when comparing results for $k$ and $\kappa$. 

We then write the shape function up to linear order in density: 
\begin{equation}\label{eq:shapelin}
S_k(\rho) = \frac{1}{2}\bigg[1-\tanh(Y_k\rho_c)  + Y_k(1-\tanh^2(Y_k\rho_c))\rho + \dots\bigg]= C_k + D_k \rho + \dots
\end{equation}
From the reasoning in the previous section, we know that the constant $\mu_k$, which sets the values of the coefficients of the TRS breaking terms, is proportional to $C_kD_k$ for a shape function of the form in (\ref{eq:shapelin}). $C_k$ and $D_k$ do not change sign for any $k$, so we conclude that the coefficients of the TRS breaking terms do not change sign either. 

The value of $C_k$ decreases monotonically as $k$ increases. However, there is a critical $k$ value at which $\partial_k D_k = 0$, which occurs at the value of $k$ for which $1=2Y_k\rho_c\tanh(Y_k\rho_c)$. We believe that this non-monotonicity may explain the non-monotonic dependence of the typical cluster size on the switching rate, $k$, that we saw in the numerical results (see the last section of the SI).

Finally, when the switching rate becomes small enough ($k\ll 10^{-2}$) the slope of our shape function may become large enough and our above argument may break down as the convergence of the infinite sums above is then not guaranteed, i.e. we may have $Y_k>\frac{\pi}{2\rho_c}$. Nevertheless, we still expect microphase separation in the low switching rate regime. Indeed, we expect that, for any $k>0$, there will be some non-zero outflow of agents from the drop (after turning from \textit{on} to \textit{off}) which will eventually balance the inflow of agents from the background fluid as a single drop grows. This would imply a finite radius at which any drop would stop growing and imply a reversal of the Ostwald Ripening process. A sufficiently large system would be able to support multiple drops of this size coexisting, which we identify as microphase separation. A similar argument is given in \cite{Zwicker2015} for the presence of microphase separation for small but non-zero chemical reaction rates.

In short, our intuition relies on the following coarse-grained model of droplet growth: consider a single liquid droplet of radius $R$ much smaller than the system size $L$, i.e. $\sigma\ll R\ll L$ and assume that the droplet is made up entirely of \textit{on} particles, such that the volume fraction of \textit{on} and \textit{off} particles in the droplet are given respectively by $\phi_l$ and 0. This approximation is particular appropriate in the low switching rate regime, where we have already shown that a high level of demixing is observed. Finally, we assume that the distribution of particles surrounding the drop is homogeneous; the relative abundance of each type of particles is thus determined by the switching rates. Let $\phi_g$ be the volume fraction of the surrounding system gas, then the volume fraction of \textit{on} and \textit{off} agents is given by $\frac{1}{2}\phi_g$ for equal switching rates.
	
The volume (area in two dimensions) of the drop $V$ is affected by: (1) an influx of \textit{on} agents attracted from the surrounding fluid through the boundary, $\partial V$, and (2) a net loss of {\it on} agent when switching \textit{off} within the drop. In two dimensions, the equation for the area of the drop can thus be written as
\begin{equation}
\frac{dV}{dt} = \frac{k_{\rm{in}}\phi_g\phi_l}{2}\partial V - k V\phi_l,
\end{equation}
where $k_{\rm{in}}$ is the rate at which \textit{on} particles are attracted to the drop from the background fluid. Assuming a spherical drop, we have $V = \pi R^2$, which allows us to write an equation for the evolution of the droplet radius as
\begin{equation}
 \frac{dR}{dt} = \frac{k_{\rm{in}}\phi_g \phi_l}{2} - \frac{k \phi_l R}{2}.
\end{equation}

In the passive case, $k=0$, we easily see that the radius of the drop grows indefinitely, at least when the size of the drop is much less than the size of the system. As the size of the drop approaches the system size, we need to consider other mechanisms which will prevent infinite growth of the drop like resource limitations. Introducing switching ($k>0$) leads to a finite steady state for the drop size in the system, given by
\begin{equation}
R_{\infty} = \frac{k_{\rm{in}}\phi_g}{k}. 
\end{equation}
This finite-valued steady state for $R$ implies that the switching dynamics are restricting the Ostwald Ripening process and further imply that multiple drops of finite size may coexist and be stable at stationarity.

\section{Numerical analysis of kinetic equations}

\subsection{Non-dimensionalization of the R-DDFT equations}
Our R-DDFT equations (\ref{seq:rddft}) are each of the form 
\begin{equation}
\partial_t \rho_i(\rv,t) = \nabla \cdot\bigg[D\nabla\rho_i(\rv) + \frac{1}{\gamma}\rho_i(\rv) \nabla\big[\mu_i(\rho(\rv))\big]\bigg] + k\rho_j(\rv) - k\rho_i(\rv), \quad i\ne j.
\end{equation}
Each of the terms in these equations have for dimension $T^{-1}L^{-2}$ (for spatial dimension $d=2$). To see this for the chemical potential term, we recognize that $\mu(\rho) = f'(\rho)$ where $f(\rho)$ is a free-energy density, with dimensions of an energy per length squared. It follows that $\mu$ has the dimensions of an energy. This is canceled by the friction term which can also be written as $\gamma = k_BT/D$ through Einstein's relation, and thus has dimensions of a length divided by an energy times a time. Note that here we have assumed unit mass.
	
We assume the same scaling as in the non-dimensionalization of the Langevin equation above:
\begin{equation}
\rv^*=\frac{\rv}{\sigma}, \quad \mu^*=\frac{\mu}{\epsilon} \quad\text{and}\quad t^* = \frac{Dt}{\sigma^2}.
\end{equation}
Making these substitutions, we derive the non-dimensional equation
\begin{equation}\begin{aligned}
\partial_{t^*}{\rho^*_i}(\rv^*, t) &= \nabla_{\rv^*}\cdot\bigg[\nabla_{\rv^*} {\rho^*_i}(\rv^*) + \frac{\epsilon}{D\gamma}\rho^*_i(\rv^*)\nabla\big[\mu^*_i\big(\rho^*(\rv^*)\big)\big]\bigg] + \kappa \rho^*_j(\rv) - \kappa \rho^*_i(\rv)\\
						      &=\nabla_{\rv^*}\cdot\bigg[\nabla_{\rv^*} {\rho^*_i}(\rv^*) + \Gamma\rho^*_i(\rv^*)\nabla\big[\mu^*_i\big(\rho^*(\rv^*)\big)\big]\bigg] + \kappa \rho^*_j(\rv) - \kappa \rho^*_i(\rv).
\end{aligned}\end{equation}
where the two dimensionless variables are exactly the same as those in the non-dimensionalization of the Langevin equation:
\begin{equation}
\Gamma = \frac{\epsilon}{\gamma D} = \frac{\epsilon}{k_BT}\quad\text{and}\quad \kappa = \frac{k\sigma^2}{D}.
\end{equation}

\subsection{Computational details}
	
We solve the non-dimensionalized kinetic equations numerically to confirm the results of our study of the microscopic model. To do so, we need to evaluate, $\mu_{\rm{rep}}(\rho)$ and $\mu_{\rm{atr}}(\rho)$, the repulsive and attractive contributions to the chemical potential, respectively. Splitting up a Lennard-Jones-like potential in this way is common in the literature and suitable approximations exists for both components. The attractive contribution is written as a mean-field approximation above which can be evaluated exactly by performing the convolution numerically. We do this by evaluating the product of the interaction potential and the density of \textit{on} agents in Fourier space before transforming back to real space. 
	
The contribution due to repulsive interactions is not something that we can evaluate exactly due to the divergent potential, so we use an approximation. Significant effort has been devoted to writing a closed form for this chemical potential \cite{Roth2010, Barrat2003}. We use the simple but accurate result from \textit{scaled particle theory} which sets
\begin{equation}
\mu_{\rm{rep}}(\eta'(\rv)) = k_BT\bigg[-\log(1-\eta') + \frac{3\eta'-2\eta'^2}{(1-\eta')^2}\bigg].
\end{equation}
Here, we have defined a rescaled packing fraction $\eta'(\rv)=0.8\eta(\rv)$, where the packing fraction $\eta(\rv)$ is itself defined in terms of the agent density as $\eta(\rv)= \frac{\pi \sigma_*^2}{4}\rho(\rv)$. A similar method was used in the study of active Brownian particles in \cite{Hermann2019}.
	
Note that as we use $\epsilon$ as the energy scale in our non-dimensionalization scheme, the non-dimensionalised chemical potential for repulsive interactions reads
\begin{equation}
\mu^*_{\rm{rep}}(\eta'(\rv)) = \frac{k_BT}{\epsilon}\bigg[-\log(1-\eta') + \frac{3\eta'-2\eta'^2}{(1-\eta')^2}\bigg]= \frac{1}{\Gamma}\bigg[-\log(1-\eta') + \frac{3\eta'-2\eta'^2}{(1-\eta')^2}\bigg]
\end{equation}
	
We can now solve the non-dimensionalized kinetic equations. We fix the total density of agents $\bar{\rho}$ and size of the solution domain and vary the switching rate $\kappa$. We set $\epsilon\gg k_bT$ as to ensure phase separation from a nearly-homogenous initial condition. We use a centered finite-difference scheme of $8^{th}$-order for spatial derivatives on a $200\times 200$ square grid with spacing $dx=0.15$ and periodic boundary conditions. We run the simulations until we judge that the system to have reached a stationary state. We use the \textit{ODE23} solver from Matlab which uses explicit time-integration with a variable time-step for computational efficiently. It is well understood that R-DDFT equations are stiff \cite{TeVrugt2020a}, thus the choice of the \textit{ODE23} solver rather than a higher order, more accurate solver is motivated by a compromise between accuracy and computational efficiency.

\subsection{Presence of reverse Ostwald ripening and microphase separation}	

\begin{figure}[th!]
	\centering
	\includegraphics[width=17.5cm]{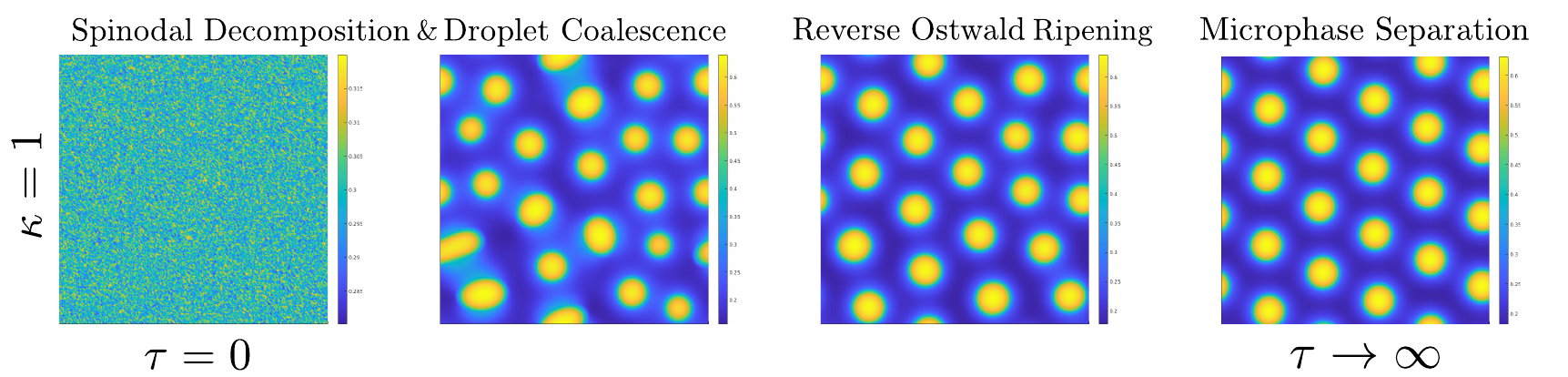}		
	\caption{\textit{Microphase Separation in Numerical Analysis of R-DDFT Equations}. We solve the kinetic equations numerically to demonstrate the presence of microphase separation at $\kappa = 1$. An initially nearly-homogeneous distribution of agents exhibits spinodal decomposition and forms a system full of mesoscopic droplets of varying sizes after some droplet coalescence. In time, the coarsening process leads to a homogeneous distribution of droplet sizes.}
	\label{sfig:mps}
\end{figure}

In Fig.\,\ref{sfig:mps}, we show examples of the time evolution of the density field in the regime where microphase separation is observed.

\subsection{Comparison of results with AMB+ deterministic phase diagram}

\begin{figure}[h]
	\centering
	\includegraphics[width=17.5cm]{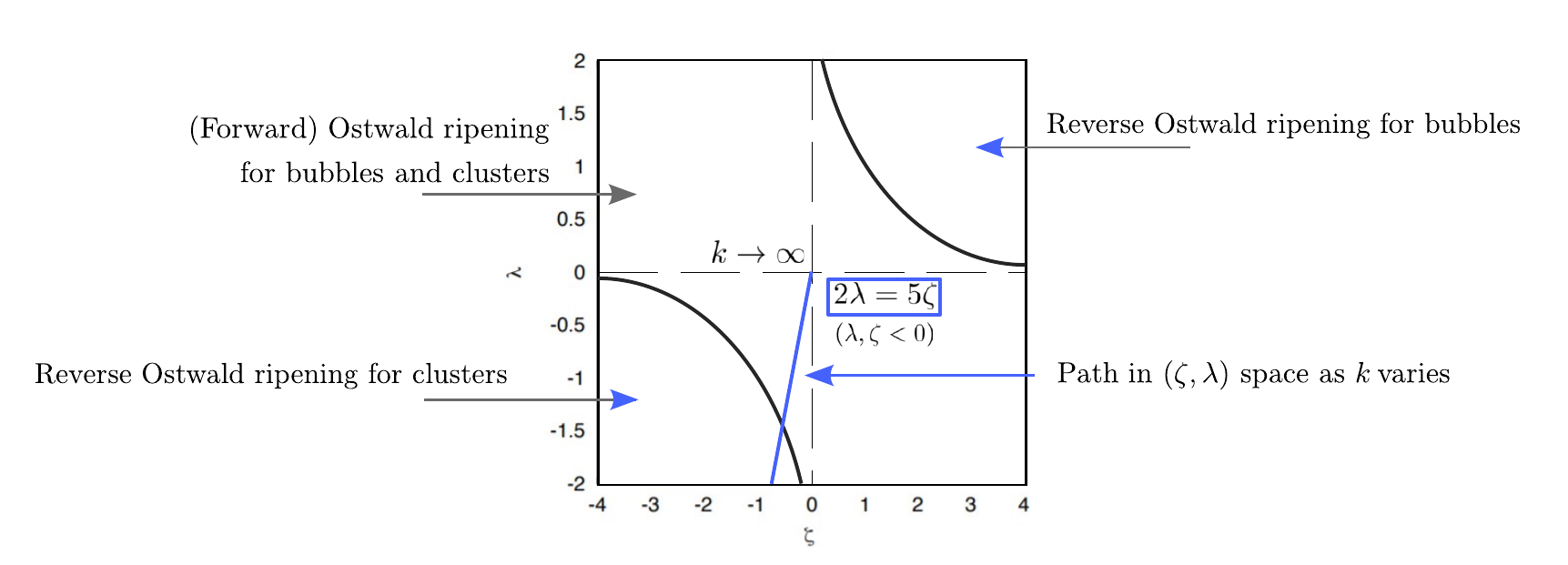}		
	\caption{\textit{Phase Diagram for Mean Field \textit{Active Model B+} equation}. We compare our results to the phase diagram for the deterministic \textit{Active Model B+} equation \cite{Tjhung2018}. There are two regions for which the equation exhibits reverse Ostwald ripening depending on the signs of the coefficients. We identify the path that the coefficients of the two time reversal symmetry breaking terms in the model of the current work take in $(\zeta, \lambda)$ space for $k\in[10^{-2},\infty)$, colored in blue. The phase diagram predicts reverse Ostwald ripening for low switching rates and Ostwald ripening for fast switching. We conclude on good agreement between the phase diagram of \cite{Tjhung2018} and the current work.}
	\label{sfig:ambpd}
\end{figure}

We comment here on the agreement between our results and those of the mean-field phase diagram for \textit{\textit{Active Model B+}} \cite{Tjhung2018}. Our formal coarse-graining procedure provided us with a relationship between the coefficients of the two time reversal symmetry breaking terms for our system, namely $2\lambda=5\zeta$. We have also shown that our constants are non-positive for all switching rates considered in the numerical analysis. We plot this curve in the $(\zeta, \lambda)$ plane in Figure \ref{sfig:ambpd}.

The results of \cite{Tjhung2018} predict that the system will exhibit microphase separation when $\lambda\zeta \gg 0$. We conclude that there is good agreement between our work and the phase diagram of \cite{Tjhung2018}: when $k \to \infty$, we argue that our system pertains to \textit{Model B} whose dynamics are entirely equilibrium. \textit{Model B} is well understood to exhibit Ostwald Ripening and hence full phase separation. We showed this analytically by arguing that $S_k(\rho)\approx\frac{1}{2}$ for high enough switching rates, thus $\lambda, \zeta\approx0$, so we are in the correct region in the phase diagram for Ostwald ripening. For fast (but finite) switching rates, our results show that the system is of AMB+ type but $\lambda$ and $\zeta$ close enough to zero (i.e. the gradient of $S_k(\rho)$ is shallow enough) that our system shows Ostwald ripening and full phase separation.

As we lower further the switching rates, we see that the gradient of $S_k(\rho)$ increases further, hence by our work above the magnitudes of the TRS breaking terms' coefficients increase and the system can exhibit microphase separation according to the above phase diagram. Indeed, we confirmed this phenomena by solving both the equations of motion of the microscopic model and the R-DDFT kinetic equations numerically. We also argued above that the coefficients of the TRS breaking terms behave non-monotonically with $k$.  Said differently, the value of $k$ governs how deep in the microphase separation regime the system is. As such, the non-monotonic behavior of $\lambda$ and $\zeta$ with $k$ could  explain the non-monotonic relation between the switching rate and the typical cluster size observed in our numerical analysis.

As detailed above, we expect the system to continue to exhibit microphase separation even for very low switching rates ($k\ll 10^{-2}$). We thus believe that the coefficients of the TRS breaking terms will remain large and negative for all positive $k$. 
	

%